\documentclass{article}
\usepackage{amsfonts,amssymb,amsmath}
\usepackage[dvips]{epsfig}
\usepackage[T1]{fontenc}
\usepackage[latin1]{inputenc}
\usepackage{graphicx}
\usepackage[english]{babel}
\usepackage{amsmath}
\usepackage{amssymb}
\usepackage{amsfonts}
\usepackage{color}

\begin{document}

\title{\bf Braneworld Black Holes and Entropy Bounds }
\author{Y. Heydarzade$^{1,2}$\thanks{%
email: heydarzade@azaruniv.edu}, H. Hadi$^{1}$\thanks{%
email: h.hadi@azaruniv.edu}, C. Corda$^{2}$\thanks{%
email: cordac.galilei@gmail.com; Corresponding author} ~and F. Darabi$^{1,2}$\thanks{%
email: f.darabi@azaruniv.edu} ,
\\{\small $^1$Department of Physics, Azarbaijan Shahid Madani University, Tabriz, Iran}\\ {\small $^2$Research Institute for Astronomy and Astrophysics of Maragha (RIAAM), Maragha 55134-441, Iran}}
\date{\today}

\maketitle
\begin{abstract}
The Bousso's $D$-bound entropy for the various possible black
hole solutions on a $4$-dimensional brane is checked. It is found that the $D$-bound entropy here is apparently different from that of
obtained for the  $4$-dimensional black hole solutions. This difference is
interpreted as the extra loss of information, associated to
the extra dimension, when an extra-dimensional black hole is moved outward  the observer's cosmological horizon. Also, it is discussed that $N$-bound entropy is hold for the possible solutions here. Finally, by adopting
the recent Bohr-like approach to black hole quantum physics for the excited
black holes,  the obtained results are written also in terms of the black hole
excited states.
\end{abstract}
\maketitle
\section{Introduction}

Black hole (BH) quantum physics started in the 70s of last century
with the remarkable works of Bekenstein \cite{key-1,key-2} and Hawking
\cite{key-3}. It is a general conviction that Hawking radiation
\cite{key-3} and Bekenstein-Hawking entropy \cite{key-2,key-3} are
the two most important provisions of a yet unknown theory of quantum
gravity which will permit to unify Einstein's general theory of relativity
(GTR) with quantum mechanics. In fact, researchers in quantum gravity
think that BHs should be the fundamental bricks of quantum gravity
in the same way that atoms are the fundamental bricks of quantum mechanics
\cite{key-4}. In this framework, a fundamental result again by Bekenstein
was that BHs have the maximum entropy for given mass and size which
is allowed by quantum theory and by the GTR \cite{key-5}. This Bekenstein
bound represents an upper limit on the entropy that can be contained
within a given finite region of space having a finite amount of energy.
In other words, it is the maximum amount of information required to
perfectly describe a given physical system down to the quantum level
\cite{key-5}. Thus, assuming that the region of space and the energy
of the system are finite, the information necessary to perfectly describe
it, is also finite \cite{key-5}. Bekenstein bound has also important
consequences in the physics of information and in computer science
because it is connected with the so-called Bremermann's Limit \cite{key-6},
a maximum information-processing rate for a physical system having
finite size and energy.  Also, the field equations of the GTR can be also
derived by assuming the correctness of the Bekenstein bound and of
the laws of thermodynamics \cite{key-7}. Today we have various arguments
which show that some form of the bound must exist if one wants the
laws of thermodynamics and the GTR to be mutually consistent \cite{key-8}.
A generalization of the Bekenstein bound was attempted by Bousso \cite{key-9},
who conjectured an entropy bound having statistical origin and to
be valid in all space-times admitted by Einstein's equation. This
is the covariant entropy bound and reduces to Bekenstein bound for
systems with limited self-gravity \cite{key-9}. Bousso also proposed
a bound on the entropy of matter systems within the cosmological horizon
called the ``$D$-bound'' which claims that the total observable entropy
is bounded by the inverse of the cosmological constant \cite{R.B}.
The
dependance of   $D$-bound  on the cosmological constant and the area of
 initial horizon  \cite{R.B} turns out to be useful for at
least one application \cite{ab}, but its relationship to the flat space
Bekenstein bound  still remains obscure. If one implies the cosmological horizon in terms of gravitational
radius rather than the energy of the system, the $D$-bound will be the same as
the Bekenstein bound. 
Note that although a special limit is taken but the agreement is non-trivial, because
the background geometry differs significantly from the flat space. So, the $D$-bound in its general form
may be regarded  as a de Sitter space equivalent  of the flat space Bekenstein
bound. Moreover, Bousso  derived the Bekenstein bound from Geroch process  to the higher
dimensional spacetimes, i.e $ D >\ 4$. It has been done  for both of the asymptotically de Sitter spaces (associated to
the cosmological horizon) and for asymptotically flat spaces (associated
to a black hole).

In this paper,  we will discuss Bousso's $D$-bound in the framework of
the braneworld BHs \cite{key-30}, see also \cite{key-11} as a review on the
braneworld
black holes. These objects arise from the gravitational
collapse of matter trapped on a brane \cite{key-12}. In particular,
Bousso's $D$-bound entropy will be derived for the various possible
BH solutions on $4D$ brane. After that, it will be shown that the
relation between the $D$-bound and Bekenstein entropy bound put restriction
on the braneworld BH solutions or gives counterexample for the $D$-bound. 

The
organization of the paper is as follows. In  section 2, we review the
entropy bounds. Then, in section 3, we introduce
the general vacuum black hole solution and its subclasses on  a $4D$ brane obtained in \cite{key-30}.
In the following sections, we investigate $D$-bound entropy for the these mentioned subclasses. The paper ends in section 9, with some concluding remarks.

\section{The Entropy Bounds}
Following Bousso \cite{R.B}  for the $D$-bound on matter
entropy in de Sitter space, one can infer some physical results by
considering the $D$-bound and its relationship with Bekenstein entropy
bound. In this line, to derive $D$-bound, one may suppose a matter system
within the apparent cosmological horizon of an observer. In such a
situation, the observer is in a universe with a future de-Sitter asymptotic.
By moving relative 
to the matter system toward the asymptotic region, this observer can be witness of a thermodynamical process by which
the matter system is moved outward  the cosmological horizon. Then,
the observer will find himself in the space-time that has been converted
to empty pure de-Sitter space. In this process, the initial thermodynamical
system, the asymptotic de Sitter space including the matter system, has
the total entropy
\begin{equation}
S=S_{m}+\frac{A_{c}}{4},\label{entropyi}
\end{equation}
where $S_{m}$ is the entropy of the matter inside the cosmological
horizon and $A_{c}/4$ is the Bekenstein-Hawking entropy associated
to the enclosing apparent cosmological horizon. At the end of the
process, the final entropy of the system will be $S_{0}=A_{0}/4$
in which $A_{0}$ is the area of the cosmological horizon of the  de Sitter space empty of any matter. Now, regarding the generalized second law, i.e. $S\leq S_{0}$,
one arrives at \cite{R.B} 
\begin{equation}
S_{m}\leqslant\frac{1}{4}(A_{0}-A_{c}).\label{dbound}
\end{equation}
This is the so-called \emph{$D$-bound} on the matter systems in an asymptotically
de Sitter space. For empty de Sitter space, we have $A_{0}=A_{c}$
and consequently the $D$-bound vanishes, because there is no matter
present. Using the fact $S_{m}\geqslant0$, one realizes $A_{c}\leqslant A_{0}$.
Then, a matter system enclosed by a cosmological horizon has smaller
area than the horizon area of an empty de-Sitter space. Now, consider
a BH, as the matter system, in an asymptotically de Sitter space.
For this case, the matter entropy $S_{m}$ is the BH's Bekenstein-Hawking
entropy. Then, one can verify that this new configuration also satisfies
the $D$-bound and consequently, the area of the cosmological horizon
surrounding the black hole $A_{c}$ will be smaller than $A_{0}$,
the cosmological horizon of the empty de Sitter space.

 The metric for this case is given by
\begin{equation}
ds^{2}=-f(r)dt^{2}+f(r)^{-1}dr^{2}+r^{2}d\Omega_{2}^{2},\label{bh}
\end{equation}
where 
\begin{equation}
f(r)=1-\frac{2M}{r}-\Lambda r^{2}.\label{fr}
\end{equation}
Setting $f(r)=0$ gives the locations of the BH horizon $r_{b}$ and
the cosmological horizon  $r_{c}$ in the presence of the BH, respectively.
The values of these horizon locations depends on the mass parameter
$M$ such that by setting $M=0$, we arrive at the empty de Sitter
space. Then, for the latter case, the only positive root of $f=0$
is $r_{c}=r_{0}$ representing the radius of the cosmological horizon.
But in the case of $M>0$, there is another root $r_{b}$ which represents
the BH horizon $(r_{b}\approx2M$ for small mass parameter $M$).
By increasing the mass parameter $M$, the BH horizon $r_{b}$ increases
while the cosmological horizon $r_{c}$ decreases. Remember that for
the empty de Sitter space, the cosmological horizon has area $A_{0}=4\pi r_{0}^{2}$
while there exists a matter inside the system, the cosmological horizon
has area $A_{c}=4\pi r_{c}^{2}$. For $M=0$, $r_{c}$ has its maximum
value as $r_{c}=r_{0}$ and decreases for $M>0$. Then, for all range
of $M$, we have $A_{c}<A_{0}$ and we can consider the $D$-bound.

The $D$-bound ($\ref{dbound}$) states that for this case, the BH entropy
i.e. $A_{b}/4=\pi r_{b}^{2}$, is less than $\pi(r_{0}^{2}-r_{c}^{2})$.
Thus, we can consider the total entropy of the system of  matter, which is enclosed by the
cosmological horizon (the entropy of Schwarzschild-de Sitter space), as 
\begin{equation}
S=\pi(r_{c}^{2}+r_{b}^{2}),\label{sm}
\end{equation}
and we know that this entropy is less than the entropy of empty de-Sitter
space 
\begin{equation}
S_{0}=\pi r_{0}^{2}.
\end{equation}
By solving the cubic equation ($\ref{fr}$), i.e. $f(r)=0,$ finding
its positive roots and putting it into ($\ref{sm}$), we can rewrite
the system's total entropy $S$ in the following form for small $M$ parameter 
\begin{equation}\label{m}
S=\pi r_{0}^{2}(1-\frac{2M}{r_{0}})+O(M^{2}).
\end{equation}
Now, we can investigate the relation between this $D$-bound with the
Bekenstein entropy bound. For the de Sitter space, the energy of the
system is not well-defined because there is no suitable asymptotic
region for this space. However, Birkhoff's theorem \cite{key-34}
implies that there exists some Schwarzschild-de Sitter solution for
a spherical system such that its metric is the same as the metric
at large radii. This large radii can be regarded as the cosmological
horizon radius, i.e $r_{c}$. Then, one can call this BH as the \emph{system's
equivalent BH}, and its radius is the gravitational radius $r_{g}$
of the system.

In the flat space, the gravitational radius is exactly twice the mass-energy,
i.e. $r_{g}=2M$ \cite{key-34}, and one can express the Bekenstein
bound in terms of both these quantities, i.e mass-energy or gravitational
radius. However, for an asymptotically de Sitter space, one can still
define the gravitational radius $r_{g}$, while the mass-energy cannot
be defined. In this case, one may define $A_{0}$ in the $D$-bound,
the relation ($\ref{dbound}$), in terms of $r_{c}$ and $r_{g}$
rather than $r_{0}$. 

Also, the mass parameter can be obtained in terms of the BH radius
by the help of the equation ($\ref{fr}$) as 
\begin{equation}\label{gogi}
2M=r_{b}\left(1-\frac{r_{b}^{2}}{r_{0}^{2}}\right).
\end{equation}
By setting $r_{b}=r_{g}$, we can express $A_{0}$ in terms of the
$r_{g}$ and $r_{c}$. For this purpose, in the limit of small \emph{
equivalent BHs} $(r_{g}\ll r_{c})$ which corresponds to light matter
systems, one finds
\begin{equation}
r_{0}^{2}=r_{c}^{2}\left(1+\frac{r_{g}}{r_{c}}\right)+O\left[(\frac{r_{g}}{r_{c}})^{2}\right].
\end{equation}
Then, by rewriting $A_{0}$ using this equation, the $D$-bound ($\ref{dbound}$),
to the first order in $r_{g}$ reads as 
\begin{equation}
S_{m}\leq\pi r_{g}r_{c}.\label{bb}
\end{equation}
 Now, recall the Bekenstein entropy bound defined in terms of gravitational
radius $r_{g}=2M$ as 
\begin{equation}
S_{m}\leq\pi r_{g}R,\label{nn}
\end{equation}
where $R$ is the radius of the sphere enclosing the system. By comparing
($\ref{bb}$) with ($\ref{nn}$), we find that the $D$-bound  coincides
with Bekenstein bound, using the fact that in de Sitter space a stable
system cannot be larger than $R=r_{c}$.

 For the excited BHs, i.e. the BHs which emitted a large amount of Hawking
quanta, the recent Bohr-like approach to BH quantum physics \cite{key-4,key-35,key-36}
permits to write the BH gravitational radius in function of the BH
quantum level as \cite{key-4,key-35,key-36} 
\begin{equation}
\left(r_{g}\right)_{n}=2M_{n}=2\sqrt{M_{I}^{2}-\frac{n}{2}},\label{eq: Mn}
\end{equation}
where $n$ is the BH principal quantum number, $M_{n}$ is the mass
of the BH excited at the level $n$ and $M_{I}$ is the initial BH
mass, i.e. the BH mass before that Hawking radiation starts to be
emitted. In fact, in \cite{key-4,key-35,key-36} the intuitive but
general conviction that BHs result in highly excited states representing
both the ``Hydrogen atom'' and the ``quasi-thermal emission''
in quantum gravity has been shown to be correct, because the Schwarzschild
BH results somewhat similar to the historical semi-classical hydrogen
atom introduced by Bohr in 1913 \cite{key-37,key-38}. For the excited
BHs, using the Eq. (\ref{eq: Mn}), the  Eqs. from (\ref{gogi}) to (\ref{nn}) become
\begin{equation}
2\sqrt{M_{I}^{2}-\frac{n}{2}}=r_{b}(1-\frac{r_{b}^{2}}{r_{0}^{2}}),
\end{equation}
\begin{equation}
r_{0}^{2}=r_{c}^{2}(1+\frac{2\sqrt{M_{I}^{2}-\frac{n}{2}}}{r_{c}})
+O\left(\frac{2M_{I}^{2}-\frac{n}{2}}{r_{c}^{2}}\right),
\end{equation}

\begin{equation}\label{gn}
S_{m}\leq2\pi r_{c}\sqrt{M_{I}^{2}-\frac{n}{2}},
\end{equation}
and finally 
\begin{equation}\label{mn}
S_{m}\leq2\pi R\sqrt{M_{I}^{2}-\frac{n}{2}}.
\end{equation}
Then, it is seen that for
the excited BHs, we have tighter bound depending on the   BH principal quantum number $n$.

Here, it is worth mentioning to a brief discussion on the $N$-bound entropy.
The $N$-bound states that the observable entropy $ S$  in any universe which has
a positive cosmological constant $ \Lambda$ is bounded by $ N= \frac{3\pi}{\Lambda}$,
regardless of its matter content \cite{ab}.
For our purpose, one can write the $N$-bound as follows
\begin{equation}\label{N}
S= S_m+S_c\leq N.
\end{equation}
Here, using (\ref{m}) and (\ref{N}), we have 
\begin{equation}
\pi r_{0}^{2}(1-\frac{2M}{r_{0}})\leq\frac{3\pi}{\Lambda}=\pi r_0^2,
\end{equation}
which indicates that the $N$-bound is also satisfied.

In the Following section, we introduce the general vacuum black hole solution and its subclasses on  a $4D$ brane.
Then, we discuss on the corresponding entropy bounds related to each of these
subclasses.%%%%%%%%%%%%%%%%%%%%%%%%%%%%%%%%%%%%%%%%%%%%%%%%%%%%%%%%%%%%%%%%%%%%%%%%%%%%%%%%%%%%%%%%%%%%%%%%%
\section{Vacuum Black Hole Solutions on the Brane}

In a braneworld model,  the visible Universe with 3 spatial dimensions
is considered as being restricted to a brane inside a higher-dimensional
space. If one assumes the additional dimensions to be compact (that
is curled up in themselves and having their lengths of order of the
Planck length), such dimensions are inevitably within the Universe. Instead,
if one assumes the additional dimensions to be not compact, the higher-dimensional
space is called \emph{the bulk.} In that case, on one hand, other
branes can move through the bulk. On the other hand, some extra dimensions
can be extensive and even infinite. A first attempt to discuss a braneworld
model was the pioneering work \cite{key-13}. More than 15 years later,
we find the works of Gogberashvili \cite{key-14},  Randall-Sundrum
(RS) scenarios, i.e. RS1 \cite{key-17} and RS2 \cite{key-18},  Arkani-Hamed-Dimopoulos-Dvali
(ADD) model \cite{Nima},  Dvali-Gabadadze-Porrati (DGP) model \cite{Dvali}, see \cite{Marteens}
and \cite{Langlois} for a review on brane gravity. In
this Section, we start from the black hole solution \cite{key-30} in the most general braneworld model introduced in \cite{Marcus} and developed in \cite{key-22}-\cite{shahram3}, without giving any detail on this model and
its applications.
 In this  general model,  there is
no specific junction conditions or $Z_{2}$ symmetry and consequently,
this model differs from the usual RS braneworld scenario where $Z_2$ symmetry is applied across a background $4D$
brane considered as a boundary embedded in an ambient bulk space.
In this case, the extrinsic curvature of the background
boundary is completely determined by the confined energy-momentum tensor
on the brane using the   Israel-Darmois-Lanczos (IDL) condition.

We consider a $4D$ brane spacetime $(\mathcal{M}_{4},g)$ embedded
in a $5D$ bulk space $(\mathcal{M}_{5},\mathcal{G})$. In order to
obtain the vacuum solution on the brane, we assume that the $4D$
brane $(\mathcal{M}_{4},g)$ is devoid of matter fields and the $5D$ ambient
bulk space $(\mathcal{M}_{5},\mathcal{G})$ possesses a constant curvature.
Then, using the Gauss-Codazzi equations \cite{key-29}, the following
induced equations on the $4D$ brane can be obtained \cite{key-30}
\begin{equation}\label{field}
G_{\mu\nu}=Q_{\mu\nu},
\end{equation}
where 
\begin{equation}
Q_{\mu\nu}=(K_{\mu}^{\,\,\,\,\,\,\gamma}K_{\gamma\nu }-KK_{\mu\nu })-\frac{1}{2}(K\circ K-K^{2})g_{\mu\nu},\label{Q}
\end{equation}
is a completely geometrical quantity resulted from the extrinsic curvature
$K_{\mu\nu}$ of the $4D$ embedded brane where also we defined the terms
$K\circ K\equiv K_{\mu\nu}K^{\mu\nu}$ and $K\equiv g^{\mu\nu}K_{\mu\nu}$.

It is clear that the right hand side of Eq. (\ref{field}) appears
as the modification to the vacuum field equations on the brane with
respect to the standard Einstein field equations of the GTR. This modification has a geometric origin and
is resulted from the extrinsic curvature of the $4D$ brane embedded
within its $nD$ ambient bulk space.
From the spirit of $Q_{\mu\nu}$, one can verify that it is an
independently conserved quantity, i.e. it possesses a null divergence as
$\nabla_{\mu}Q^{\mu\nu}=0$. 

In order to find a general static spherically symmetric BH spacetimes,
one can consider the following metric 
\begin{equation}
ds^{2}=-e^{\mu(r)}dt^{2}+e^{\nu(r)}dr^{2}+r^{2}(d\theta^{2}+sin^{2}(\phi)d^{2}\phi).
\end{equation}
Then, using the Codazzi equation, the induced field equations (1)
and the conservation equation $\nabla_{\mu}Q^{\mu\nu}=0$, one can
obtain the non-vanishing components of the extrinsic curvature tensor
$K_{\mu\nu}$ as 
\begin{equation}
K_{00}(r)=-\alpha e^{\mu(r)},\label{ex}
\end{equation}
\begin{equation}
K_{11}(r)=\alpha e^{\nu(r)},
\end{equation}
\begin{equation}
K_{22}(r)=\alpha r^{2}+\beta r,
\end{equation}
\begin{equation}
K_{33}(r,\theta)=\alpha r^{2}\sin^{2}{\theta}+\beta r\sin^{2}{\theta},
\end{equation}
where $\alpha$ and $\beta$ are integration constants. Using the
components of the extrinsic curvature tensor, one can find the non-vanishing
components of the $Q_{\mu\nu}$ tensor as 
\begin{eqnarray}
Q_{00} & = & -\frac{g_{00}}{r^{2}}\left(3\alpha^{2}r^{2}+4\alpha\beta r+\beta^{2}\right),\nonumber \\
Q_{11} & = & -\frac{g_{11}}{r^{2}}\left(3\alpha^{2}r^{2}+4\alpha\beta r+\beta^{2}\right),\nonumber \\
Q_{22} & = & \frac{g_{22}}{r}\left(-3\alpha^{2}r-2\alpha\beta\right),\nonumber \\
Q_{33} & = & \frac{g_{33}}{r}\left(-3\alpha^{2}r-2\alpha\beta\right).
\end{eqnarray}
Then, by solving the field equations ($\ref{field}$), one can find
the metric components as 
\begin{equation}
e^{\mu(r)}=e^{-\nu(r)}=1-\frac{2M}{r}-\alpha^{2}r^{2}-2\alpha\beta r-\beta^{2},\label{metric}
\end{equation}
where $M$ is the central BH mass and $\alpha$ and $\beta$ are integration
constants resulted from the extrinsic curvature of the embedded brane
which are playing the role of cosmological parameters. In this regard,
the three modification terms in ($\ref{metric}$), relative to the
familiar Schwarzschild solution \cite{key-31}, have geometric origins
and are arising from the non-trivial extrinsic geometry of the brane
within its higher dimensional ambient bulk space. Then, regarding
the metric functions ($\ref{metric}$), one can distinguish the following
distinct subclasses: 
\begin{itemize}
\item The case of $\beta=0$,
\item The case of $\alpha^{2}\simeq0$, 
\item The case of $\alpha\neq0$ and $\beta\neq0$, 
\item The case of $\alpha^{2}\simeq0$ and $\beta^{2}\simeq0$, 
\item The case of $\alpha=\beta=0$, 
\item The case of $\alpha=0$ and $\beta\neq0$, 
\item The case of $M=0$,\item The case of $M=\beta=0$, 
\item The case of $M=\alpha=0$.
\end{itemize}
In the following sections, we investigate the $D$-bound and its relation
with the Bekenstein entropy bound for each of these distinct solutions
on brane in detail. 

\section{The Entropy Bounds for the Case of $\beta=0$}
This solution corresponds to the metric functions 
\begin{equation}
e^{\mu(r)}=e^{-\nu(r)}=1-\frac{2M}{r}-\alpha^{2}r^{2},
\end{equation}
representing the Schwarzschild-de Sitter BH with positive cosmological
constant, i.e. $\Lambda=\alpha^{2}$ \cite{key-32}. Interestingly, in this case,
the cosmological constant has a geometric origin, rather than its
ad-hoc introduction to the field equations of  GR and arises
from the extrinsic curvature of the brane in a higher dimensional
bulk. Then, the discussions here on the entropy bounds are the same as the
section 2  and we avoid to repeat. The only point is that there is
a geometric origin for the cosmological constant of the de-Sitter space.   \section{The Entropy Bounds for the Case of $\alpha^{2}\simeq0$ }

The corresponding solution is given by 
\begin{equation}
e^{\mu(r)}=e^{-\nu(r)}=1-\frac{2M}{r}-2\alpha\beta r-\beta^{2}.
\end{equation}
Except the $\beta^{2}$ term, this solution looks like to the Kiselev
BH \cite{key-39} in GR, see also its generalization to the Rastall theory
\cite{rastall}. In this solution,  the BH is surrounded by a quintessence
field with the field structure parameter $\sigma=2\alpha\beta$ \cite{key-39}.
Then, this can be called as the Schwarzschild-quintessence-like BH
on the brane. Now, let us find the solution of the equation 
\begin{equation}
f(r)=1-\frac{2M}{r}-2\alpha\beta r-\beta^{2}=0.\label{f}
\end{equation}
For $M=0$ or equivalently in the BH absence, the Eq. (\ref{f}) gives
\begin{equation}
r_{0}=\frac{1-\beta^{2}}{2\alpha\beta},\label{r0}
\end{equation}
which represents the cosmological horizon of an empty of matter (BH)
space. In the presence of BH , for $\alpha\beta>0$ and $1-\beta^{2}>0$,
there are two solutions for the Eq. $(\ref{f}$) as
\begin{equation}
r_{c}=\frac{(1-\beta^{2})+\sqrt{(\beta^{2}-1)^{2}-16\alpha\beta M}}{4\alpha\beta},\label{rC}
\end{equation}
and 
\begin{equation}
r_{b}=\frac{(1-\beta^{2})-\sqrt{(\beta^{2}-1)^{2}-16\alpha\beta M}}{4\alpha\beta}.\label{rB}
\end{equation}
In the case of excited BHs, one can use the Eq. (\ref{eq: Mn}) and rewrite
these solutions in terms of the BH quantum level and of the BH initial
mass as 
\begin{equation}
r_{c}=\frac{(1-\beta^{2})+\sqrt{(\beta^{2}-1)^{2}-16\alpha\beta\sqrt{M_{I}^{2}-\frac{n}{2}}}}{4\alpha\beta},\label{rCn}
\end{equation}
and 
\begin{equation}
r_{b}=\frac{(1-\beta^{2})-\sqrt{(\beta^{2}-1)^{2}-16\alpha\beta\sqrt{M_{I}^{2}-\frac{n}{2}}}}{4\alpha\beta}.\label{rBn}
\end{equation}
Then, for the next conveniences, we rewrites $r_{c}^{2}$ as 
\begin{equation}
r_{c}^{2}=\frac{(1-\beta^{2})r_{b}}{2\alpha\beta}\left(1-\frac{2M}{1-\beta^{2}}\right),\label{rzero}
\end{equation}
which becomes 
\begin{equation}
r_{c}^{2}=\frac{(1-\beta^{2})r_{b}}{2\alpha\beta}\left(1-\frac{2\sqrt{M_{I}^{2}-\frac{n}{2}}}{1-\beta^{2}}\right),\label{rzeron}
\end{equation}
for the excited BHs. 

Interestingly, from the Eqs. ($\ref{rC}$) and ($\ref{rB}$), we find $r_{b}+r_{c}=r_{0}$.
Now, let us consider the $D$-bound and its relationship with Bekenstein
bound. To check the $D$-bound, suppose a matter system (BH) within the
apparent cosmological horizon of an observer. This observer lies in
a Universe which is going to be asymptotically quintessence-like Universe
in the future. The observer can be witness of a thermodynamical process
by which the matter system (BH) is dropped across the cosmological
horizon. Then, he will be in the space-time that has been converted
to empty quintessence-like space with the radius $r_{0}$. In this
process, the initial thermodynamical system has entropy given by the Eq.
($\ref{entropyi}$). In this way, $S_{m}$ is the entropy of the matter
(BH) inside the cosmological horizon and $A_{c}$ is the area of cosmological
horizon in the presence of matter system (BH). At the end of process,
the final entropy of the system will be $S_{0}=A_{0}/4$, using the
fact that the quarter of the area of the apparent cosmological horizon
is the Bekenstein-Hawking entropy. Here, $A_{0}$ is the area of the horizon
of the empty quintessence-like space. Then, using the generalized
second law ($S\leq S_{0}$) leading to the $D$-bound of the Eq. ($\ref{dbound}$),
we find 

\begin{equation}
S_{m}\leqslant\pi\left(\left(\frac{1-\beta^{2}}{2\alpha\beta}\right)^{2}-\left(\frac{1-\beta^{2}}{2\alpha\beta}\right)r_{c}+\frac{Mr_{c}}{\alpha\beta}\right).\label{Nbulk}
\end{equation}
Using the Eq. ($\ref{r0}$), one can rewrites the Eq. (\ref{Nbulk}) as 
\begin{equation}
S_{m}\leqslant\pi\left((\frac{1-\beta^{2}}{2\alpha\beta})(r_{0}-r_{c})+\frac{Mr_{c}}{\alpha\beta}\right).\label{SM}
\end{equation}
 On the other hand, taking the limit of $r_{b}$ for small $\beta$
values, one gets 
\begin{equation}
r_{b}=\frac{2M}{1-\beta^{2}}.\label{eq: 36}
\end{equation}
 Now, if one recalls the approach in the section 2, one can take $r_{b}=r_{g}$.
Then, we have 
\begin{equation}
2M=r_{g}(1-\beta^{2}).\label{eq: 37}
\end{equation}
Putting the Eq. (\ref{eq: 37}) and  Eq. (\ref{r0}) in the Eq. ($\ref{SM}$) and using the relation
$r_{b}+r_{c}=r_{0}$, one obtains
\begin{equation}\label{S_m}
S_{m}\leqslant\pi r_{g}r_{c}\left(1+r_{c}\right) +r_g^2(1+r_c).
\end{equation}

Here, it is worth to discuss about the terms appeared in the RHS of
(\ref{S_m}) in comparison to the term in the RHS of (\ref{bb}).
Actually, it turns out that the  terms in the RHS of
(\ref{S_m}) are appeared because of the extra loss of information, associated to
the extra dimension, when an extra-dimensional black hole is moved outward  the observer's cosmological horizon.  

 For further discussion we compare this with the covariant entropy bound of a 4-dimensional black
hole. The  covariant entropy bound for this black hole leads to \cite{key-9}
\begin{equation}
S_m\leqslant \frac{A}{4},
\end{equation}
where $A$ is the area of black hole's horizon. So,  here the covariant entropy
bound becomes $S_m\leqslant \pi r_g^2$. This bound is tighter than the bound
in inequality (\ref{S_m}). This result is not surprising because even for
 a 4-dimensional black
hole, the covariant entropy bound is tighter than $D$-bound \cite{R.B}.  
    
Regarding the $N$-bound, we
note that since   $\alpha^{2}\simeq0$ which means $ \Lambda\simeq 0$, then  the $N$-bound becomes infinite and  so this kind  of black hole in a braneworld may be allowed.

\section{The Entropy Bounds for the Case of $\alpha\protect\neq0$ and $\beta\protect\neq0$}

This is the most general solution on the brane where 
\begin{equation}
\begin{array}{c}
f(r)=1-\frac{2M}{r}-\alpha^{2}r^{2}-2\alpha\beta r-\beta^{2}\\
\\
=1-\frac{2M}{r}-(\alpha r+\beta)^{2}.
\end{array}\label{eq: 42}
\end{equation}
 In the BH absence, i.e $M=0$, the Eq. (\ref{eq: 42}) gives 
\begin{equation}
(\alpha r_{0}+\beta)^{2}=1,\label{c}
\end{equation}
where $r_{0}$ is the radius of the cosmological horizon in the absence
of any matter system (BH). Using the Eq. ($\ref{c}$), one can rewrite
the Eq. $(\ref{eq: 42}$) as 
\begin{equation}
f(r)=1-\frac{2M}{r}-\frac{(\alpha r+\beta)^{2}}{(\alpha r_{0}+\beta)^{2}}.\label{sigma}
\end{equation}
Now, by solving $f(r)=0$, one can find two roots $r_{c}$ and $r_{b}$
as the cosmological horizon in the presence of matter system (BH)
and the BH horizon, respectively, as 
\begin{equation}
1-\frac{2M}{r_{c}}-\frac{(\alpha r_{c}+\beta)^{2}}{(\alpha r_{0}+\beta)^{2}}=0,\label{hfh}
\end{equation}
and 
\begin{equation}
1-\frac{2M}{r_{b}}-\frac{(\alpha r_{b}+\beta)^{2}}{(\alpha r_{0}+\beta)^{2}}=0.\label{gg}
\end{equation}
If $2M\ll r_{c}$, using $(\ref{hfh}$), we have 
\begin{equation}
(\alpha r_{0}+\beta)^{2}=(\alpha r_{c}+\beta)^{2}(1+\frac{2M}{r_{c}}+\frac{4M^{2}}{r_{c}^{2}})+O(\frac{2M}{r_{c}})^{3},\label{dVt1}
\end{equation}
which leads to 
\begin{equation}
r_{0}^{2}=r_{c}^{2}+4M^{2}+\frac{2\beta}{\alpha}(r_{c}-r_{0})+2Mr_{c}+\frac{2M\beta^{2}}{r_{c}\alpha^{2}}+\frac{4M\beta}{\alpha}.\label{eq: rzerozero}
\end{equation}
 Also, we can rewrite the equation ($\ref{gg}$) as 
\begin{equation}
r_{b}=2M\left(1-(\frac{\alpha r_{b}+\beta}{\alpha r_{0}+\beta})^{2}\right)^{-1}.\label{eq: r zero}
\end{equation}\label{r_b}
We note that the condition $2M\ll r_{c}$ will be, in principle, satisfied
for the excited astrophysics BHs in the future, when a lot of their mass
will be radiated in terms of Hawking quanta. In that case, if one
uses again the Eq. (\ref{eq: Mn}), the Eqs. from (\ref{dVt1}) to (\ref{eq: r zero})
can be re-written in terms of the BH quantum level and  the BH initial
mass as 
\begin{eqnarray}\label{dVt1n}
(\alpha r_{0}+\beta)^{2}&=&(\alpha r_{c}+\beta)^{2}(1+\frac{2\sqrt{M_{I}^{2}-\frac{n}{2}}}{r_{c}}
+\frac{4\left(M_{I}^{2}-\frac{n}{2}\right)}{r_{c}^{2}})\nonumber\\
&&+O(\frac{2\sqrt{M_{I}^{2}-\frac{n}{2}}}{r_{c}})^{3},
\end{eqnarray}
and
\begin{eqnarray}\label{eq: rzerozeron}
r_{0}^{2}&=&r_{c}^{2}+4\left(M_{I}^{2}-\frac{n}{2}\right)+\frac{2\beta}{\alpha}(r_{c}-r_{0})\nonumber\\
&&+2\sqrt{M_{I}^{2}-\frac{n}{2}}r_{c}+\frac{2\sqrt{M_{I}^{2}-\frac{n}{2}}\beta^{2}}{r_{c}\alpha^{2}}+\frac{4\sqrt{M_{I}^{2}-\frac{n}{2}}\beta}{\alpha},
\end{eqnarray}
and 
\begin{equation}
r_{b}=2\sqrt{M_{I}^{2}-\frac{n}{2}}\left(1-(\frac{\alpha r_{b}+\beta}{\alpha r_{0}+\beta})^{2}\right)^{-1}.\label{eq: r zeron}
\end{equation}
Now, we can investigate the relation between the $D$-bound and the Bekenstein
bound for this type of BHs on the brane. To derive the $D$-bound, suppose a
matter system within the apparent cosmological horizon of an observer.
Regarding the Eq. ($\ref{eq: 42}$), although the structure of the whole
system is different than the usual Schwarzschild-de Sitter system,
particularly for the mean distances, we find that here the observer
is also in a Universe which is going to be asymptotically de Sitter
in the future. The observer can be witness of a thermodynamical process
by which the matter system is dropped across the cosmological horizon.
Then,  he will be in the space-time that has been converted to empty
de Sitter-like space but not pure de Sitter. In this process the initial
thermodynamical system has entropy, like as the equation ($\ref{dbound}$),
where $S_{m}$ is the entropy of the matter inside the cosmological
horizon and $A_{c}$ is the area of the cosmological horizon. A quarter
of the area of the apparent cosmological horizon is the Bekenstein-Hawking
entropy. At the end of process, the final entropy of the system will
be $S_{0}=A_{0}/4$. Here, $A_{0}$ is the area of horizon of the empty
de Sitter-like space. Then, the generalized second law, $S\leq S_{0}$,
leads to $D$-bound 
\begin{equation}
S_{m}\leqslant2\pi M\left(2M+r_{c}+\frac{\beta^{2}}{r_{c}\alpha^{2}}+\frac{2\beta}{\alpha}\right)+\frac{2\beta}{\alpha}(r_{c}-r_{0}),\label{eq: Nbulk}
\end{equation}
where we used the Eq. ($\ref{dbound}$). Using the Eq. ($\ref{rzero}$), we
can rewrite this relation as 
\begin{equation}
S_{m}\leqslant\pi r_{b}\left(1-(\frac{\alpha r_{b}+\beta}{\alpha r_{0}+\beta})^{2}\right)\left(r_{b}\left(1-(\frac{\alpha r_{b}+\beta}{\alpha r_{0}+\beta})^{2}\right)+r_{c}+\frac{\beta^{2}}{r_{c}\alpha^{2}}+\frac{2\beta}{\alpha}\right)+\frac{2\beta}{\alpha}(r_{c}-r_{0}).\label{}
\end{equation}
Replacing $r_{b}=r_{g}$, one gets 
\begin{equation}
S_{m}\leqslant\pi r_{g}\left(1-(\frac{\alpha r_{g}+\beta}{\alpha r_{0}+\beta})^{2}\right)\left(r_{g}\left(1-(\frac{\alpha r_{g}+\beta}{\alpha r_{0}+\beta})^{2}\right)+r_{c}+\frac{\beta^{2}}{r_{c}\alpha^{2}}+\frac{2\beta}{\alpha}\right)+\frac{2\beta}{\alpha}(r_{c}-r_{0}).\label{eq: 55}
\end{equation}
which yields 
\begin{eqnarray}
S_{m} & \leqslant & \pi r_{g}r_{c}\nonumber \\
 &  & +\pi r_{g}\left(r_{g}\left(1-(\frac{\alpha r_{g}+\beta}{\alpha r_{0}+\beta})^{2}\right)+\frac{\beta^{2}}{r_{c}\alpha^{2}}+\frac{2\beta}{\alpha}\right)\nonumber \\
 &  & -\pi r_{g}\left(\frac{\alpha r_{g}+\beta}{\alpha r_{0}+\beta}\right)^{2}\left(r_{g}\left(1-(\frac{\alpha r_{g}+\beta}{\alpha r_{0}+\beta})^{2}\right)+r_{c}+\frac{\beta^{2}}{r_{c}\alpha^{2}}+\frac{2\beta}{\alpha}\right)\nonumber \\
 &  & +\frac{2\beta}{\alpha}(r_{c}-r_{0}).
\end{eqnarray}
By comparing this $D$-bound with the Bekenstein entropy bound ($\ref{nn}$)
with $r_{g}=r_b \cong2M$ for $r_b\ll r_0$, using the equation (\ref{r_b}) and $r_{c}=R$, we find that there are two physical possibilities
as \begin{itemize}
\item The extra terms should vanish in order to maintain the Bekenstein
bound for this type of black holes on
the brane. 
\item The extra terms should possesses total negative values, but small
relative to $\pi r_{c}r_{g}$, in order to lead a $D$-bound tighter
than the Bekenstein bound for this type of black holes on the brane. 
\end{itemize}
Both of these two possibilities put restrictions on the geometric
parameters $\alpha$ and $\beta$ of the embedded brane within its
ambient space.

If one sets $\beta=0$, one finds
\begin{equation}
S_{m}\leqslant\pi r_{c}r_{g}(1+\frac{r_{g}}{r_{c}}),\label{Komar}
\end{equation}
where in the limit of $r_{g}\ll r_{c}$, the relation $S_{m}\leqslant\pi r_{c}r_{g}$
in ($\ref{bb}$) can be recovered,  as the obtained result in \cite{R.B}. 

Also in this case, if one considers excited BHs, the equations from (\ref{eq: Nbulk}) to (\ref{Komar}) can be written in terms of the BH quantum level
and the BH initial mass through the Eq. (\ref{eq: Mn}). For this case,
the Eq. (\ref{Komar}) becomes 
\begin{equation}\label{Komar n}
S_{m}\leqslant\pi r_{c}r_{g}\left(1+\frac{2\sqrt{M_{I}^{2}-\frac{n}{2}}}{r_{c}}\right).
\end{equation}
Here, it is also seen that for
the excited BH, we have tighter entropy bound relative to the initial their
states.

We can discuss here about the $N$-bound  for the solution  (\ref{eq: 42}). Because  the cosmological constant term $\alpha^2 r^2$ in the spacetime of the metric (\ref{eq: 42}) is still asymptotically dominant term, so it
is expected that the $N$-bound will hold also for the solution (\ref{eq: 42}) similar to the de-Sitter and Schwarzschild-de-Sitter spaces.
\section{The Entropy Bounds for the Case of $\alpha^{2}\simeq0$ and $\beta^{2}\simeq0$}

The corresponding solution is given by

\begin{equation}
e^{\mu(r)}=e^{-\nu(r)}=1-\frac{2M}{r}-2\alpha\beta r,
\end{equation}
which is exactly the Schwarzschild BH in the quintessence field \cite{key-39}
with the quintessence structure parameter $\sigma=2\alpha\beta$.
Now, we need to find the solutions of 
\begin{equation}\label{eq:*60}
f(r)=1-\frac{2M}{r}-2\alpha\beta r-\beta^{2}=0.
\end{equation}
 For $M=0$ or equivalently in the BH absence, the Eq. (\ref{eq:*60})
gives 
\begin{equation}
r_{0}=\frac{1}{2\alpha\beta},
\end{equation}
which represents the cosmological horizon of an empty of matter (BH)
space. In the presence of BH, there are two solutions for $(\ref{eq:*60}$)
as 
\begin{equation}
r_{c}=\frac{1+\sqrt{1-16\alpha\beta M}}{4\alpha\beta},\label{rCD}
\end{equation}
and 
\begin{equation}
r_{b}=\frac{1-\sqrt{1-16\alpha\beta M}}{4\alpha\beta},\label{rB1}
\end{equation}
representing the cosmological horizon in the BH presence and the BH horizon,
respectively. Here, it is useful to rewrite ($\ref{rCD}$) as
\begin{equation}
r_{c}^{2}=\frac{1}{2\alpha\beta}(r_{c}+2M).
\end{equation}
Then, by using $r_{b}+r_{c}=r_{0}$ and the generalized second law
($S\leq S_{0}$), we find the $D$-bound 
\begin{eqnarray}
S_{m} & \leq & \pi r_{0}^{2}-\pi r_{c}^{2}\nonumber \\
 & = & \pi((r_{c}+r_{g})^{2}-r_{0}(r_{c}+2M))\nonumber \\
 & = & \pi(r_{0}r_{g}-2r_{0}M)\nonumber \\
 & = & \pi(2r_{c}M+r_{g}^{2}-2r_{0}M).\label{eq: 65}
\end{eqnarray}
Since $r_{0}\gg r_{g}$ then the third term overcomes to the second,
i.e. $r_{g}^{2}-2r_{0}M<0$, and this relation represents a tighter
bound than the Bekenstein and covariant entropy bounds for this type of BHs on the brane.
Then, these BHs can  exist as the real physical BH solutions on the
brane, if one regards the covariant bound as the basic physical entropy
bound.

For this case, if one considers the excited BHs, the Eq. (\ref{eq: Mn}) permits to rewrite
the Eq. (\ref{eq: 65}) in terms of the BH excited state as 

\begin{eqnarray}
S_{m} & \leq & \pi r_{0}^{2}-\pi r_{c}^{2}\nonumber \\
 & = & \pi\left((r_{c}+r_{g})^{2}-r_{0}(r_{c}+2\sqrt{M_{I}^{2}-\frac{n}{2}})\right)\nonumber \\
 & = & \pi\left(r_{0}r_{g}-2r_{0}\sqrt{M_{I}^{2}-\frac{n}{2}}\right)\nonumber \\
 & = & \pi\left(2r_{c}\sqrt{M_{I}^{2}-\frac{n}{2}}+r_{g}^{2}-2r_{0}\sqrt{M_{I}^{2}-\frac{n}{2}}\right),\label{eq: 66}
\end{eqnarray}
representing a tighter bound. 

 Because of  $\alpha^{2}\simeq0$ which means $ \Lambda\simeq 0$, the $N$-bound becomes infinite and  so this kind  of black hole in a braneworld may be allowed.

\section{The Remaining Cases}
\subsection{The  Case of $\alpha=\beta=0$.}

Here, the corresponding metric is the  familiar Schwarzschild solution
\cite{key-34}

\begin{equation}
e^{\mu(r)}=e^{-\nu(r)}=1-\frac{2M}{r}.
\end{equation}
In this case, there is no cosmological horizon. Thus, one cannot consider
the thermodynamical process defined  for obtaining the $D$-bound or $N$-bound.
Therefore,
that
method cannot be applied for this case. \textcolor[rgb]{0.501961,0,1}{}
\subsection{The  Case of $\alpha=0$ and $\beta\protect\neq0$.}

In this case, we have no cosmological horizon and consequently, we
can not define the thermodynamic process considered in \cite{R.B} and
\cite{ab} to obtain the
$D$-bound or $N$-bound. 

\subsection{The  Case of $M=0$.}

In this case, we have two horizons without BHs. If the inner horizon
can be go out of the outer horizon, like the BH in the thermodynamical
process in the studied  cases in sections 3-6, it may be possible to find $D$-bound or $N$-bound.
This is an issue that will be analysed in our future work \cite{key-40}. \textcolor[rgb]{0.501961,0,1}{}

\subsection{The  Case of $M=\beta=0$}

This case represents the pure de Sitter space where $A_{c}=A_{0}$
and the $D$-bound vanishes. \textcolor[rgb]{0.501961,0,1}{}

\subsection{The  Case of $M=\alpha=0$}

There is no cosmological horizon for this case to define the mentioned
thermodynamical process in \cite{R.B}. \textcolor[rgb]{0.501961,0,1}{}

\section{Concluding Remarks}

In this paper, we have focused on the Bousso's $D$-bound entropy and on the Bekenstein's
entropy bound. In particular, Bousso's $D$-bound entropy has been  checked
for the various possible extra dimensional black hole solutions.
It turns out that the $D$-bound entropy here is apparently different from that of
obtained for the  $4$-dimensional black hole solutions. This difference is
interpreted as the extra loss of information, associated to
the extra dimension, when an extra-dimensional black hole is moved outward  the observer's cosmological horizon. 
We have also discussed briefly about the $N$-bound entropy for the possible black hole solutions
on the braneworld, represented by the cases $\alpha^2=0$ and $\alpha^2\ne0$. It turns
out that the N-Bound holds for both cases. In addition, through the recent Bohr-like approach to black hole quantum physics
for the excited black holes, it has been possible to rewrite the various obtained
results
also in function of the black hole quantum principal number, i.e. in function
of the black hole quantum excited state.
In this regard, we have tighter entropy bound for the excited black holes relative their initial states.
 We hope to further extend our analysis
in a future paper \cite{key-40}. 

\textcolor[rgb]{1,0,0.501961}{ }

\section{Acknowledgements}

This paper has been supported financially by the Research Institute
for Astronomy and Astrophysics of Maragha (RIAAM), Project Number
No. 1/4717-41.

\end{document}